\documentclass[useAMS,usenatbib]{mn2e}
\usepackage{epsfig}
\usepackage{natbib}
\usepackage{subfigure}
\usepackage{multirow}
\usepackage{amssymb}

\newcommand{\kms}{\mbox{km s$^{-1}$}}
\newcommand{\msun}{\mbox{M$_{\odot}$}}

\title[Predicting the Properties of Merger Remnants]
{Predicting the Properties of the Remnants of Dissipative Galaxy Mergers}

\author[Covington et al.]
{M. Covington$^{1,4}$\footnotemark[1]\footnotemark[2], A. Dekel$^{1,2}$, T. J. Cox$^3$,  P. Jonsson$^4$, and J.~R. Primack$^{1,4}$ \\
        $^1$Department of Physics, University of California, 1156 High St., Santa Cruz, 95064, USA\\
	$^2$Racah Institute of Physics, The Hebrew University, Jerusalem, 91904, Israel\\
        $^3$Harvard-Smithsonian Center for Astrophysics, 60 Garden St., 
		Cambridge, MA 02138, USA\\
	$^4$Santa Cruz Institute of Particle Physics, University of California,
		1156 High St., Santa Cruz, 95064, USA\\	
	}

\begin{document}
\maketitle
\begin{abstract}
We construct a physically motivated model for predicting the properties
of the remnants of gaseous galaxy mergers, given the properties of the
progenitors and the orbit.  The model is calibrated using a large suite 
of SPH merger simulations.  It implements generalized energy conservation 
while accounting for dissipative energy losses and star formation.  
The dissipative effects are evaluated from the initial gas fractions 
and from the orbital parameters via an ``impulse" parameter, which
characterizes the strength of the encounter.  Given the progenitor 
properties, the model predicts the remnant stellar mass, half-mass 
radius, and velocity dispersion to an accuracy of 25\%.  
The model is valid for both major and minor mergers.
We provide an explicit recipe for semi-analytic models of galaxy formation.
\end{abstract}
	\footnotetext[1]{National Science Foundation Graduate Fellow} 
	\footnotetext[2]{email: mdcovin@physics.ucsc.edu}

\section{Introduction}
\label{sec:intro}

Major mergers between galaxies are central to the formation and
evolution of elliptical galaxies \citep{TT72,T77,MH94dsc}. 
The hierarchical buildup of
galaxies in the $\Lambda$CDM cosmology consists of a sequence of mergers, 
of which a significant fraction are ``major," involving progenitors with
a mass ratio larger than 1:3. The gravitational interactions in such mergers 
have a dramatic effect on the dynamics and morphology of the galaxies,
in particular turning rotating disks into pressure-supported spheroids.  
If the progenitors also contain gas, the mergers induce starbursts 
followed by gas consumption, which leads to aging stellar populations.
The modeling of major mergers is therefore a key element in the attempts 
to confront the broad picture of galaxy formation with detailed observations. 
This is commonly performed via simulations incorporating Semi-Analytic Models 
(SAMs), where the complex physical processes are modeled using simplified 
parametric recipes.

Advanced SAMs are currently attempting the non-trivial task of following 
the sizes and internal velocities of galaxies. For disk galaxies, sizes are 
evaluated using the halo virial radii $R_{\rm vir}$ and spin parameters 
$\lambda$ via $R_{\rm disk} \simeq \lambda R_{\rm vir}$, with some 
modifications due to the halo density profile (Fall \& Efstathiou 1980; 
Mo, Mao \& White; Bullock, Dekel et al. 2001; Dutton et al. 2006).
The sizes of the remnants of gas-poor (``dry") mergers, where the dominant
interaction is gravitational, can be extracted from the properties of the
progenitors and the orbital energy by assuming conservation of energy 
and relaxation to virial equilibrium \citep{Cole00}. 
These considerations work well in simulations of relatively
dry mergers, quite independently of the details of the orbit.

While dry mergers may dominate in the formation of the most massive galaxies
\citep{Naab06}, the most common mergers are ``wet" mergers of gaseous galaxies.
It is thought that gas processes play an important role in the formation of 
ellipticals \citep{RobertsonFP, Dekel06, Ciotti07}.  As demonstrated below, the
sizes predicted by dissipationless energy conservation can 
be off by a factor of a few for gas-rich mergers. Our goal is to construct a 
more accurate
recipe to predict the size and velocity dispersion of the remnant of a wet 
merger given the properties of the progenitors and the orbital parameters.
Cosmological mergers involve a complex mixture of variables
(such as orbital parameters, gas fractions, mass ratio and bulge fraction)
and physical processes (such as star formation and feedback),
all of which can influence the properties of the merger remnants.
Given a rich suite of high-resolution SPH merger simulations 
\citep{thesis, Cox05},
that span the available parameter space and physical processes, albeit in 
a rather sparse and nonuniform manner, we seek a model that will properly 
represent the simulation results and enable an interpolation between them 
as well as an extrapolation to outside the simulated regime.
 
For such a recipe to be successful, it should be based on a toy model 
that grasps the essence of the main physical processes involved in wet 
mergers. Our intuition is guided by the finding from the simulations 
that the remnants are more compact when the initial gas fraction is higher
and when the first passage involves a stronger tidal impulse, namely
when a larger fraction of the orbital energy turns into internal kinetic
energy. 
At a first glance, this may seem surprising, as a system that gains energy 
is not expected to become more tightly bound, and indeed, the remnants of 
dry mergers are not very sensitive to the strength of the impulse.
This dependence on gas fraction and on the impulse implies that a 
higher gas fraction and a stronger impulse are associated with a higher degree
of dissipation, via shocks, collisions of gas clouds, and induced gas 
flows toward the centres of the merging systems.  The resultant higher gas 
densities enhance the energy losses to radiation, leaving behind a more 
tightly bound remnant.  In parallel, the higher degree of dissipation 
yields a stronger burst of star formation, which tends to be focused in the
central region of the remnant.
This understanding is the basis for our proposed recipe, which characterizes
the merger by the impulse at first passage, evaluates the associated
degree of dissipation and the resultant radiative energy losses and 
star formation, and accounts for these energy losses in the  
energy balance. A few free parameters with values of order unity can hopefully
compensate for the crude approximations made. These approximations include, 
for example, an assumption of structural homology between the progenitors 
and remnant.
The physically motivated recipe is then calibrated using the merger 
simulations, and its success is to be judged by its accuracy in matching 
the simulated remnant properties. 

In \S \ref{sec:methods} we describe the simulations used for this study.  
In \S \ref{sec:model} we present the details of our model for predicting 
remnant properties.
In \S \ref{sec:Mass} we generalize the model to unequal mass mergers.
In \S \ref{sec:Caveats} we discuss certain limitations of our
model, and in \S \ref{sec:Conclusions} we summarize our conclusions.
Appendix A discusses the details of our impulse approximation.
Appendix B presents an explicit recipe for SAMs.


\section{Merger Simulations}
\label{sec:methods}

\subsection{Numerical Code}
\label{sec:sims}

The numerical simulations analyzed in this work are part
of a large suite of galaxy merger simulations designed to
study the induced star formation \citep{thesis, Cox05} and
observable counterparts \citep{Sunrise, Jonsson06} of such
events.  Details of these simulations can be found in the
above references, but we include here a brief description
for completeness.

All numerical simulations performed in this work use the 
N-Body/SPH code GADGET \citep*{SpGad}.  Hydrodynamics are
included via the Lagrangian technique of smoothed particle
hydrodynamics (SPH).  We use the  ``conservative entropy''
version of SPH \citep{SpEnt}.  Gas is assumed, for simplicity, to be a 
primordial plasma that can radiatively cool via atomic and
free-free emission.

All of the numerical simulations presented here 
include star formation.  Stars are formed in regions of 
gas which are above a critical density for star formation at
a rate proportional to the local gas density and inversely
proportional to the local dynamical time-scale.  The
efficiency of star formation is fixed by requiring star
formation to follow the observed correlation between gas
and star-formation rate surface densities \citep{Kenn98}.

We also include a simple prescription to simulate the 
effects of feedback from massive stars.  This feedback
acts to pressurize the interstellar medium and regulates
the conversion of gas to stars.  Details of this model
and the parameter choices can be found in \citet{Cox05}.
Specifically, most simulations studied in this paper 
used the $n2med$ parameter set.  Under these assumptions
the gas pressure increases as the density squared; i.e,
star-forming gas has a ``stiff'' equation of state.  Other 
cases are discussed in \S \ref{sec:Caveats}.

The simulations presented here adopt a gravitational softening length $h=400$~pc
for the dark matter particles and 100~pc for the stellar and gas particles.
We remind the reader that, in GADGET, forces between neighboring particles
become non-Newtonian for separations $<2.3\,h$.

\subsection{Initial Galaxies}
\label{ssec:galmodel}

All of the simulations presented here are mergers
between two identical disk galaxies, except for cases discussed in \S \ref{sec:Mass}.  
The disk galaxy models are motivated by observations of low-redshift galaxies. 
In some cases we made systematic studies of varying progenitor galaxy
properties (e.g. varying gas fraction in the G3 gas fraction series).  While 
many of these varied cases would not look like typical low-redshift galaxies,
we made no attempt to vary properties in such a way as to 
capture variation with redshift.  Furthermore, the simulations
are not cosmological since the two galaxies are isolated.  Disk galaxies 
are constructed in equilibrium and contain dark matter, an exponential
stellar disk, an extended exponential gas disk, and some contain a 
dense central bulge.  Our suite consists of five main types of models, 
detailed in Table \ref{tab:gals}:
\begin{enumerate}
\item Z galaxies are gas-poor bulgeless disks and are roughly modeled after 
  the Milky Way.
\item D galaxies are $\frac{1}{100}$ of the mass of the Z's, are
  bulgeless disks, and have a high gas fraction.
\item Y galaxies are $\frac{1}{10}$ of the mass of the Z's, are 
  bulgeless disks, and have a high gas fraction.
\item Sbc galaxies are modeled after local Sbc-type spirals, with a small
  bulge and high gas fraction.
\item G galaxies span a range of mass, bulge fraction, and gas fraction. Their
  properties are taken from statistical samples of local galaxies, 
  including the SDSS.
  Their dark matter halos have not been adiabatically contracted \citep{Blumenthal1984, MMW},
  unlike all the other models.  
\end{enumerate}

The ratio of the gas to stellar exponential radii varies with model type.  For
the Z, D, and Y models, the gas and stellar radii are equal.  For the Sbc and
G models, the gas radii are three times the stellar radii.  For more detail 
on these models see \citet{thesis}.

\begin{table}
\begin{center}
\caption{Properties of progenitor galaxy models. $M_{\rm tot}$ is total mass, 
baryons plus dark matter; $c$ is concentration ($R_{\rm vir}/r_{\rm s}$); $M_{\rm stars}$ 
is the 
initial stellar mass; $B/D$ is the bulge-to-disk ratio; $f_{\rm g}$ is the initial
gas mass
divided by $M_{\rm tot}$; $R_{\rm 1/2}$ is the initial three-dimensional stellar half mass radius.}
\begin{tabular}{p{0.7cm}p{1.2cm}p{0.3cm}p{1.2cm}p{0.6cm}p{0.6cm}p{0.7cm}}
\hline
  Type & $M_{\rm tot}$ & $c$ & $M_{\rm stars}$ & $B/D$ & $f_{\rm g}$ & $R_{\rm 1/2}$ \\
   & ($10^{10} \msun$) &  & ($10^{10} \msun$) & & & (kpc) \\
\hline
\hline
\multicolumn{6}{l}{\underline{Milky Way Series}} \\
\hline
D & 1.4 & 20 & 0.036 & 0 & 0.025 & 1.16  \\
Y & 14.0 & 15 & 0.3 & 0 & 0.02 & 2.85  \\
Z & 143.0 & 12 & 5.1 & 0 & 0.004 & 4.04  \\
\hline
\multicolumn{6}{l}{\underline{Sbc Series}} \\
\hline
Sbc & 81.4 & 11 & 4.92 & 0.26 & 0.066 & 7.15  \\
\hline
\multicolumn{6}{l}{\underline{G Series}} \\
\hline
G0 & 5.0 & 14 & 0.1 & 0.02 & 0.012 & 1.84  \\
G1 & 20.0 & 12 & 0.5 & 0.06 & 0.010 & 2.33  \\
G2 & 51.0 & 9 & 1.5 & 0.11 & 0.009 & 2.90  \\
G3 & 116.0 & 6 & 5.0 & 0.22 & 0.011 & 3.90  \\
\hline 
\multicolumn{6}{l}{\underline{G3 Gas Fraction Series}} \\
\hline
G3gf1 & 116.0 & 6 & 3.6 & 0.32 & 0.023 & 3.49 \\
G3gf2 & 116.0 & 6 & 2.6 & 0.52 & 0.031 & 2.89 \\
G3gf3 & 116.0 & 6 & 1.5 & 1.34 & 0.040 & 1.77 \\
G3gf4 & 116.0 & 6 & 5.3 & 0.20 & 0.005 & 3.96 \\
\hline
\end{tabular}
\label{tab:gals}
\end{center}
\end{table}

\subsection{Merger Orbits}
\label{ssec:orbits}

The sizes of merger remnants are affected by the initial orbits and 
orientations.  To understand this relationship
a sufficient exploration of the merger orbit and
orientation parameter space is required.  To
this end, we perform mergers on an identical orbit
with various orientations of the merging galaxies.  We also perform mergers
with many different orbits.  Of all of the galaxy
models, the Sbc models have the largest variety of orbits
and orientations.  The majority of the orbits in the suite
are parabolic or near parabolic with eccentricities of $0.9$ to $1.0$.  
While these orbits are generally motivated by statistics from
N-body simulations \citep{Benson05, KandB06}, the distribution of orbits in 
the merger suite was not designed to exactly duplicate these statistics.

\begin{table*}
\begin{center}
\caption{The orbital parameters of each of the simulations used in this study.
Values are calculated from the orbital initial conditions assuming a 
point mass orbit. $R_{peri}$ denotes pericentric distance. $e$ denotes
eccentricity. For equal mass mergers $\theta_1$ and $\theta_2$ denote the orientation of the
first and second galaxy with respect to the orbital plane, where for $\theta_1=0$ the first galaxy is aligned with the orbital plane.  For unequal
mass mergers $\theta$ denotes the orientation of the orbit of the smaller
progenitor with respect to the orientation of the larger progenitor.  Again,
$\theta=0$ represents a case where the merger orbit aligns with the larger progenitor.}
\begin{tabular}{lccccclcccc}
\hline
\multicolumn{3}{l}{\underline{Major Mergers (55 runs)}} \\
\hline
\hline
name & $R_{peri}~(kpc)$ & $e$&$\theta_1$&$\theta_2$&&name&$R_{peri}~(kpc)$ & $e$&$\theta_1$&$\theta_2$ \\
\hline
D1mf-u1 & 1.1 & 1.0 & 0 & 30 && Sbc201-u4 & 11.0 & 1.0& 0 & 30 \\
D2mf-u1 & 2.4 & 1.0 &  0 & 30 && Sbc202-u4 & 11.0 & 1.0& 180 & 30\\
D3mf-u1 & 4.1 & 1.0 &  0 & 30 && Sbc203-u4 & 11.0 & 1.0& 180 & 210\\
D4mf-u1 & 6.4 & 1.0 &  0 & 30 && Sbc204-u4 & 5.5 & 1.0& 0 & 30\\
D5mf-u1 & 8.9 & 1.0 & 0 & 30 & & Sbc205-u4 & 44.0 & 1.0& 0 & 30\\
D6mf-u1 & 12.9 & 1.0 & 0 & 30 & & Sbc206-u4 & 11.0 & 1.0& 90 & 30\\
D7mf-u1 & 0.6 & 1.0 & 0 & 30 & & Sbc207-u4 & 11.0 & 1.0& 270 & 30\\
G0G0a-u1 & 2.24 & 0.95 & -30 & 30 & & Sbc208-u4 & 5.5 & 1.0& 180 & 30\\
G0G0r-u1 & 2.24 & 0.95 & 150 & 30 & & Sbc209-u4 & 5.5 & 1.0& 180 & 210\\
G0G0-u1 & 2.24 & 1.0 & -30 & 30 & & Sbc211-u4 & 44.0 & 1.0& 180 & 210\\
G1G1a-u1 & 2.96 & 0.95 & -30 & 30 && Sbc212-u4 & 11.0 & 0.9& 0 & 30\\
G1G1r-u1 & 2.96 & 0.95 & 150 & 30 & & Sbc213-u4 & 25.0 & 0.8& 0 & 30\\
G2G2r-u1&3.82&0.95& 150 & 30 &&Sbc214-u4&44.0&0.8& 0 & 30\\
G2G2-u1&3.82&0.95& -30 & 30 &&Sbc215-u4&100.0&1.0& 0 & 30\\
G3blv5G3blv5-u1 & 13.6 & 0.95 & -30 & 30 & & Sbc216-u4&100.0&0.8& 0 & 30\\
G3G3a-u1 & 13.6 & 0.95 & -30 & 30 &&Sbc217-u4&11.0&1.0& 90 & 90\\
G3G3b-u1&13.6 & 0.95 & -30 & 30 &&Sbc218-u4&11.0&0.9& 180 & 210\\
G3G3r-u1 & 13.6 & 0.95& 150 & 30 &&Z2m-u1&7.1&1.0& 0 & 30 \\
G3gf1G3gf1b-u1 & 13.6 & 0.95 & -30 & 30 && Z7m-u1 & 21.4 & 0.9& 0 & 30 \\
G3gf2G3gf2b-u1 & 13.6 & 0.95 & -30 & 30 && Z8m-u1 & 35.7 & 0.8& 0 & 30 \\
G3gf3G3gf3b-u1 & 13.6 & 0.95 & -30 & 30 && Z9m-u1 & 1.7 & 1.0& 0 & 30 \\
G3gf4G3gf4b-u1 & 13.6 & 0.95 & -30 & 30 && Z10m-u1 & 3.9 & 1.0& 0 & 30 \\
Y1mf-u1&2.9&1.0& 0 & 30 &&Z11m-u1&14.2&1.0& 0 & 30 \\
Y2mf-u1&5.7&1.0& 0 & 30 &&Z12m-u1&22.2&1.0& 0 & 30 \\
Y3mf-u1&10.0&1.0& 0 & 30 &&Z13m-u1&30.4&1.0& 0 & 30 \\
Y4mf-u1&15.7&1.0& 0 & 30 &&Z14m-u1&44.7&1.0& 0 & 30 \\
Y5mf-u1&21.4&1.0& 0 & 30 &&Z15m-u1&8.1&1.0& 0 & 30 \\
Y6mf-u1&31.4&1.0& 0 & 30 &\\
\end{tabular}
\begin{tabular}{lccccclcccc}
\hline
\multicolumn{3}{l}{\underline{Minor Mergers (22 runs}} \\
\hline
\hline
name & $R_{peri}~(kpc)$ & $e$&$\theta$&Mass Ratio &&name&$R_{peri}~(kpc)$ & $e$&$\theta$& Mass Ratio \\
\hline
G1G0-u3 & 2.96 & 0.95 & -30 & 1:3.9 && G3G1f-u1 & 27.2 &  0.95 & -30 & 1:5.8 \\
G1G0r-u3 & 2.96 & 0.95 & 150 & 1:3.9&& G3G1g-u1 & 64.4 & 0.95  & -30 &1:5.8 \\
G2G0-u3 & 3.82 & 0.95 & -30 & 1:10 &&  G3G1h-u1 & 120 & 0 & -30 & 1:5.8 \\
G2G0r-u3 & 3.82 & 0.95 & 150 & 1:10&& G3G2a-u1 & 13.6 & 0.95 & 0 & 1:2.3\\
G2G1-u3 & 3.82 & 0.95 & -30 & 1:2.6 &&  G3G2b-u1 & 13.6 & 0.95 & -90 & 1:2.3 \\
G2G1r-u3 & 3.82 & 0.95 & 150 & 1:2.6 && G3G2c-u1 & 13.6 & 0.95 & -60 & 1:2.3\\
G3G1a-u1 & 13.6 & 0.95 & 0 & 1:5.8 && G3G2d-u1 & 13.6 & 0.95 & 180 & 1:2.3\\
G3G1b-u1 & 13.6  & 0.95 & -90 & 1:5.8 && G3G2e-u1 & 6.8 & 0.95 & -30 & 1:2.3\\
G3G1c-u1 & 13.6  & 0.95 & -60 & 1:5.8 && G3G2f-u1 & 27.2 & 0.95 & -30 & 1:2.3\\
G3G1d-u1 &  13.6 & 0.95 & 180 & 1:5.8 && G3G2g-u1 & 64.4 & 0.95 & -30 & 1:2.3\\
G3G1e-u1 & 6.8 & 0.95 & -30 & 1:5.8 && G3G2h-u1 & 120 & 0 & -30 & 1:2.3\\

\hline
\end{tabular}
\end{center}
\end{table*}

\section{Modeling Properties of the Remnants}
\label{sec:model}

\subsection{Non-dissipative Energy Conservation Model}
\label{ssec:simple}

Energy conservation during the merger may be a useful constraint to impose.
If the progenitors and the remnant are homologous, then energy 
conservation and the virial theorem may be applied to the stellar systems 
as if they are self-gravitating without introducing a large error. 
Current SAMs employ such considerations to predict merger remnant radii
\citep{Cole00, Galics03}.  We start by summarizing the model of Cole et 
al. 

  In order to take orbital energy into account, this model 
assumes that the baryonic components of the progenitors spiral in 
under dynamical friction, losing energy to the outer dark matter halo, until
reaching a distance that equals the sum of their three-dimensional stellar half-mass radii.  
Energy conservation is assumed from this point on.
Thus the orbital energy term in the 
conservation equation is equal to the energy of a zero eccentricity (circular)
orbit of the two galaxies with a constant separation equal to 
the sum of their half-mass radii,
\begin{equation}
\label{eq:colemod}
 \frac{(M_{\rm 1} + M_{\rm 2})^2}{R_{\rm f}} = \frac{M_{\rm 1}^2}{R_{\rm 1}} + \frac{M_{\rm 2} ^2}{R_{\rm 2}} + 
 \frac{1}{c}\frac{M_{\rm 1} M_{\rm 2}}{R_{\rm 1} + R_{\rm 2}}.
\end{equation}
This equation 
relates the stellar three-dimensional half-mass radius of the remnant, $R_{\rm f}$, to the 
masses and three-dimensional stellar half-mass radii of the progenitors, $M_{\rm 1}$, $M_{\rm 2}$, $R_{\rm 1}$ 
and $R_{\rm 2}$, respectively.
For major mergers, the masses are assumed to include the total stellar mass
plus, as a rather arbitrary choice, twice the dark matter masses within 
$R_{\rm 1}$ and $R_{\rm 2}$ respectively. 
The constant $c$ is a 
structural parameter which relates $G M^2 / R$ to the actual internal
binding energy.  It is assumed to take the same value, $c=0.5$, for both the
progenitors and the remnant.

\begin{figure}
\begin{center}
\resizebox{8.0cm}{!}{\includegraphics{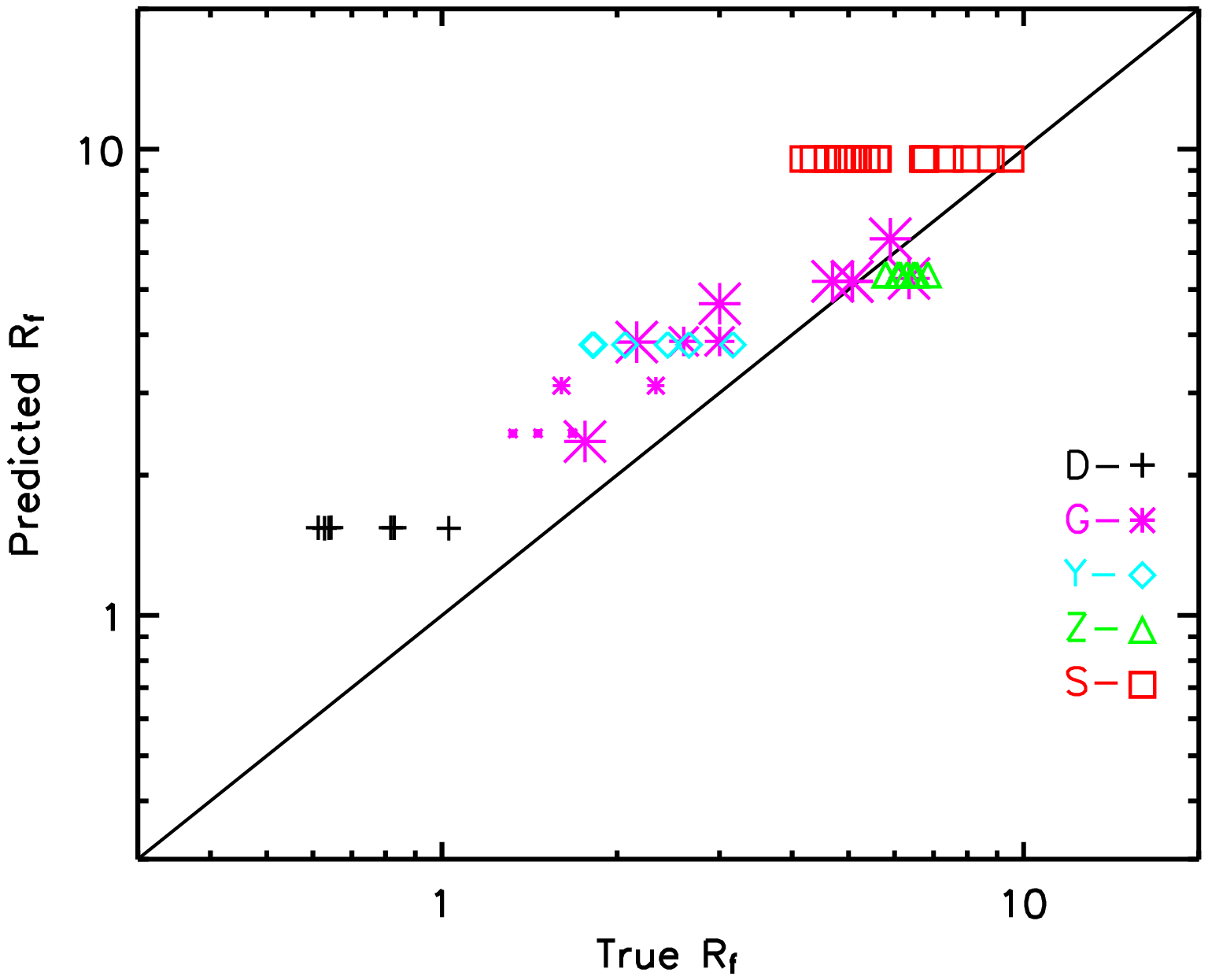}}\\
\caption{Dissipationless energy conservation model.  
The three-dimensional half-mass radii predicted by the model versus the 
actual radii in the simulated remnants.
The symbols and colors denote the type of progenitors (Table 1): 
D (black crosses), G (purple asterisks), Y (blue diamonds), Z (green
triangles) and Sbc (red squares). The symbols for G's have four different
sizes which represent the different types of G progenitors.  Larger symbols
represent more massive galaxies.  
With a relatively large scatter of $S=0.5$,
this dissipationless model systematically over-predicts the sizes of the 
remnants, because it does not address the radiative energy losses.}
\label{fig:colemod}
\end{center}
\end{figure}

This model, though crude, provides a general framework for estimating
the outcome of a given merger.  However, it has not been previously tested 
against realistic simulations that include gas dynamics and star
formation.  We begin by applying this recipe to the cases in our 
merger simulation suite.  The predictions of the model are plotted against the
actual half-mass radii of the simulated remnants in Figure~\ref{fig:colemod}.
Throughout this paper we use the fractional rms scatter, $S$, to assess 
goodness of fit.

\begin{equation}
\label{eq:chi} 
S = \sqrt{\frac{1}{N} \sum \frac{(P_{\rm predicted} - P_{\rm true})^2}{P_{\rm true}^2} },
\end{equation}
where $P$ is some property that we are trying to predict from the initial 
conditions and are measuring from the simulation for comparison.  

The predictions of this simple model deviate from the true radii by 
$S=0.50$.  Figure \ref{fig:colemod} shows that the model systematically 
over-predicts the radii of the remnants.  This is a straightforward result of 
ignoring the radiative energy losses.  Note that the predictions are best 
for the cases where the progenitors have the lowest gas fraction, cases Z and 
G3.  For each type of progenitor, we see a wide spread in actual sizes 
for a given predicted size. This results from the variations in orbits 
and orientations, which are not addressed by the model.  
Despite the shortcomings of the dissipationless energy conservation model, 
it is a useful starting point.  Our next goal is to correct for the dissipative
effects.  

\subsection{A Toy Model for Radiative Losses}
\label{ssec:new}

While the above dissipationless model works quite well in the case of
low gas fraction, dissipative losses are likely to play an important role
in the gas-rich mergers that were especially frequent in the 
early epochs of galaxy formation. In particular, they seem to be a crucial
element in the formation of the Fundamental Plane of elliptical galaxies
\citep{RobertsonFP, Dekel06}. We therefore wish to incorporate the
radiative losses in our model.

When two gas-rich galaxies merge, a number of processes cause
gas interactions and result in radiative energy losses.  Tidal 
torques during a close pass can decrease the angular momentum in the gas disks
and induce inflows into the galaxy centres. 
In a nearly radial encounter, the gas disk of one galaxy  
collides with that of the other creating shocks, which can result in 
loss of angular momentum.  Furthermore, tidal forces during a merger 
introduce orbital crossings and density perturbations within the gas disks 
which ultimately lead to an
increased gas collision rate.  As the gas clouds collide and radiate 
away their kinetic energies, they fall toward the centre of the galactic 
potential well.  This results in higher gas densities which lead to 
star formation and further radiative losses. Hence, the orbit, energy 
losses, and star formation are intimately linked.

\begin{figure}
\begin{center}
\resizebox{8.0cm}{!}{\includegraphics{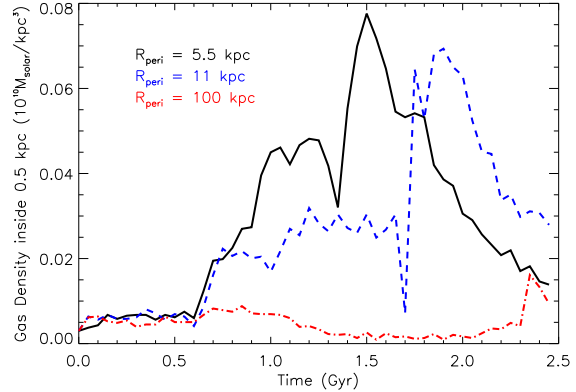}}\\
\caption{The central gas density in the inner $0.5$ kpc during three separate
Sbc mergers with initial orbital pericentric distances of 5.5, 11 and
100 kpc (solid black, dashed blue, and dash-dot red respectively).  
In all cases, the initial pass occurs at 0.6 Gyr.  
Final coalescences occur at roughly 1.5, 1.8 and 2.4 Gyr respectively.
The plot demonstrates that the central gas density in the period
following the first pass is increasing with decreasing pericentre distance, 
and that the density at the final coalescence is following the same trend.
}
\label{fig:rho_peri}
\end{center}
\end{figure}

The first step toward predicting energy losses from a given initial
configuration is to characterize the perturbative strength
of a given encounter.
More specifically, we need to assess the dependence of the gas 
collisions and dissipation on the properties of the orbit.
We find that the dissipation resulting from a given orbit can be characterized 
using the orbital parameters at the first close pass.  This may seem surprising
at first, since often most of the stars are formed in the final coalescence 
rather than at the initial encounter, but it can be explained by the sequence 
of events following the first pass.  The gas disks are perturbed during the
first pass, inducing a continuous gas in-fall toward the galactic centre,
lasting $\sim 1~{\rm Gyr}$.  A stronger disturbance in the first pass
leads to a larger buildup of gas in the progenitor centres. 
Some of this gas is involved in a first starburst immediately following
the first pass, but a large fraction of this gas serves as a reservoir for
star formation during the later stages of the merger, especially the violent
final coalescence.  Consequently, an orbit that is more disruptive on the 
first pass also suffers more energy losses, and 
forms more stars, during the final coalescence.
This is illustrated in Figure \ref{fig:rho_peri}.

During a close pass of two galaxies, orbital energy is injected into the 
internal kinetic energies of both galaxies.  We define this ``impulse'' 
as the difference between the peak in the total internal kinetic energies during the
encounter and the total initial internal kinetic energies of the two progenitors.
We approximate the impulse on galaxy 1 by
\begin{equation}
\Delta E = \frac{A G^2 M_{\rm 1,tot}^2 M_{\rm 2,tot}}{V_{\rm peri}^2 (R_{\rm peri}^2 
+ B~R_{\rm 1,tot}~R_{\rm peri} + C~R_{\rm 1,tot}^2)},
\end{equation}
where $M_{\rm 1,tot}$ and $M_{\rm 2,tot}$ are the total mass of the perturbed and 
perturbing galaxies respectively, baryons plus dark matter,
and $R_{\rm 1,tot}$ is the total half mass radius of the perturbed galaxy.  
$R_{\rm peri}$ is the ``pericentric distance'' of the first passage, as 
calculated from the initial orbit assuming point masses.  
The best-fitting values of the parameters are found to be 
$A=1.6$, $B=1.0$ and $C=0.006$.
The fit of the equation to the impulse measured from the 
simulations in shown in Figure \ref{fig:impulse}.  A more detailed discussion 
of this impulse approximation is given in Appendix A.

\begin{figure}
\begin{center}
\resizebox{8.0cm}{!}{\includegraphics{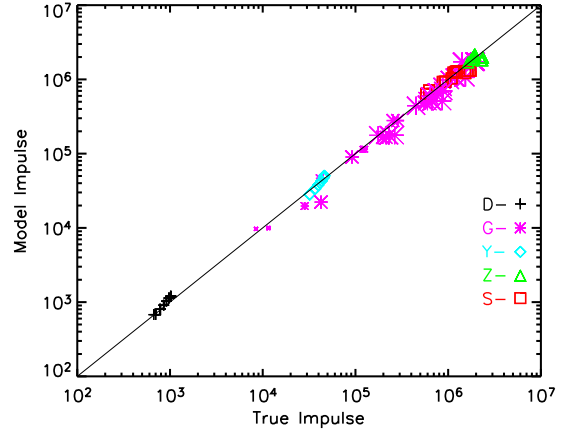}}\\
\caption{The approximation for the impulse against the measured impulse 
from the simulations. 
$S=0.40.$}
\label{fig:impulse}
\end{center}
\end{figure}

\begin{figure}
\begin{center}
\resizebox{8.0cm}{!}{\includegraphics{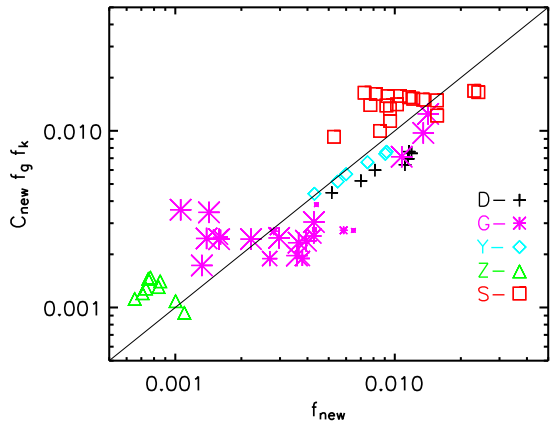}}\\
\caption{Fraction of new stars formed during the merger: the approximation
based on gas fraction and impulse versus the actual value in the simulations.
A stronger perturbation from a larger impulse enhances the collision
rate between gas clouds, and thus induces more star formation. 
}
\label{fig:sf_plot}
\end{center}
\end{figure}

We characterize the dissipative strength of a merger using the fractional impulse at the 
first pass, $f_k \equiv \Delta E/K_{\rm tot}$, where $\Delta E$ is the
impulse and $K_{\rm tot}$ is the total initial internal kinetic energy 
of the galaxy, baryons and dark matter.
We find that  $f_{\rm new}$, the fraction of new stars formed in the merger relative to
$M_{\rm tot}$, is indeed 
proportional to the fractional impulse and to the gas fraction $f_{\rm g}$,
\begin{equation}
\label{eq:sf_imp}
f_{\rm new} = C_{\rm new}~f_{\rm g}~f_{\rm k.}
\label{eq:fnew}
\end{equation}
The best fit to our simulations is obtained for $C_{\rm new}\sim 0.3$.  
The gas fraction determines the 
overall normalization of star formation, whereas the impulse factor tracks 
the variations with orbit.  
This prediction for the fraction of new stars is plotted in Figure 
\ref{fig:sf_plot} against the actual fraction of new stars in the simulated
remnants.
In order to consistently treat the star formation
in simulations that were run for different lengths of time, 
and in keeping with previous work \citep{SPF}, we separate star formation into
a merger-induced burst component and a quiescent component.  
We accomplish this by subtracting out star formation measured from 
simulations of each progenitor in isolation.  The quantity $f_{\rm new}$
in equation \ref{eq:fnew} refers to the burst component only.

Given the above correlation between impulse and star formation, we proceed
to predict the energy loss by assuming that the colliding gas 
that forms new stars during the merger loses a constant fraction of its 
kinetic energy in the
collision. In more detail, we make the following assumptions:

\begin{enumerate}
  \item Gas clouds have initial velocities approximately equal to the
    average initial velocities of the dark matter/baryon system.
  \item The average impulse per mass imparted to gas that will form new
    stars is approximately equal to the average impulse per mass for the 
    entire dark matter/baryon system.    
  \item The energy lost during the merger is proportional to the kinetic
    energy of the gas that will form stars.  This includes:
    1) the initial kinetic energy, and 2) the kinetic energy gained from 
    the impulse.
\end{enumerate}

\begin{figure}
\begin{center}
\resizebox{8.0cm}{!}{\includegraphics{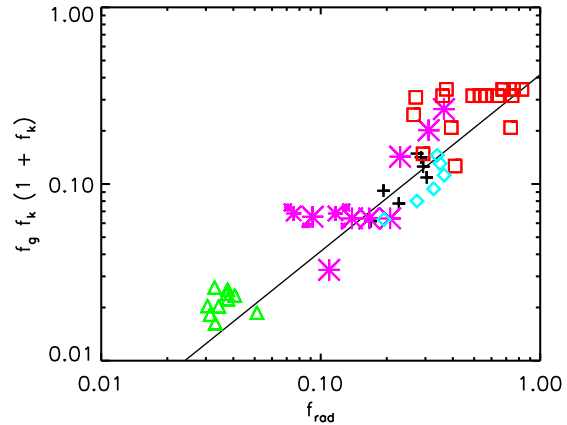}}\\
\caption{Fractional energy loss.
The approximation of radiative losses in equation \ref{eq:rad_mod2}, which is based on gas fraction
and impulse, is shown against the actual radiative energy losses. 
It demonstrates a crude proportionality, with the line plotted $y=0.4x$.
}
\label{fig:rad_mod2}
\end{center}
\end{figure}

While simplistic, these assumptions allow us to make a connection 
between radiative losses and star formation.  Assumption (i) would be 
valid if the system resembled an isothermal sphere.
Assumption (iii) provides a sensible way for parameterizing 
the energy loss, given that it is associated with collisions of gas
clouds. Under this assumption, the radiative energy loss can be written as 
\begin{equation}
E_{\rm rad} \propto (f_{\rm new}~K_{\rm tot} + f_{\rm new}~\Delta E).
\end{equation}
The first term is the total initial internal kinetic energy of the gas
that forms stars, and the second term is the energy imparted to 
that gas by the impulse.  Dividing by $K_{tot}$, and using 
equation \ref{eq:sf_imp} gives 
\begin{equation}
\label{eq:rad_mod2}
f_{\rm rad} \propto f_{\rm new}(1+f_{\rm k}) \propto ~f_{\rm g}~ f_{\rm k}~(1+f_{\rm k}).
\end{equation}
This expression is plotted in Figure \ref{fig:rad_mod2} against the 
actual radiative energy losses in our simulations, showing a crude
proportionality, with the proportionality constant $\sim 0.4$.

\subsection{A Dissipative Model for Remnant Radii}
\label{ssec:diss_model}

Guided by this toy model for dissipative losses during a galaxy merger, we
construct a modified energy conservation equation for predicting remnant radii.
Note that the final stellar mass can be computed 
using the initial mass and the equation for star formation (\ref{eq:sf_imp}). 
We choose to include in our energy formula the initial stellar mass plus 
the mass of the stars formed during the merger, such that the dissipative 
term tracks the loss of energy of the progenitor's gas that
becomes stars in the remnant.  The energy equation is
\begin{equation}
\label{eq:tot_e}
E_{\rm int,f}= E_{\rm int,i} + E_{\rm rad} + E_{\rm orb},
\end{equation}
where the final and initial internal energies are 
\begin{equation}
\label{eq:int_e_f}
E_{\rm int,f}= -\frac{C_{\rm int} G (M_{\rm 1} + M_{\rm 2} + M_{\rm new,1} + M_{\rm new,2})^2}{R_{\rm f}}\\
\end{equation}
\begin{equation}
\label{eq:int_e_i}
E_{\rm int,i}= -C_{\rm int} G\left[\frac{(M_{\rm 1} + M_{\rm new,1})^2}{R_{\rm 1}} + \frac{(M_{\rm 2} + M_{\rm new,2})^2}{R_{\rm 2}}\right],
\end{equation}
and $M_{\rm 1}$, $M_{\rm 2}$, $R_{\rm 1}$, and $R_{\rm 2}$ are the initial stellar masses and 
the corresponding three-dimensional half-mass radii of the progenitors,  
$R_{\rm f}$ is the final three-dimensional half-mass radius of the remnant, and 
$M_{\rm new,1}$ and $M_{\rm new,2}$ are the mass of new stars formed during the merger 
in galaxies one and two respectively, as predicted by our model.  
The constant structural parameter, 
which relates the internal energy of the systems to $GM^2/R$,
is determined by best fit to the radii of the simulated remnants
to be $C_{\rm int} \simeq 0.5$.  In assuming homology, we set the structural
constants for the progenitors and remnants to the same value.

The radiated energy term is the sum of the losses from the two progenitors,
\begin{equation}
\label{eq:rad_e}
E_{\rm rad} =-C_{\rm rad} \Sigma_{i=1}^2 K_{\it i} f_{{\rm g,}{\it i}} f_{{\rm k,}{\it i}} (1+f_{{\rm k,}{\it i}}), 
\end{equation}
where $K_i$, $f_{{\rm g,}{\it i}}$ and $f_{{\rm k,}{\it i}}$ are the initial internal kinetic energy,
gas fraction and impulse corresponding to progenitor $i$.
The constant of proportionality 
relating our expression for energy loss to the actual 
energy lost is determined by best fit to the radii of the simulated remnants
to be $C_{\rm rad} \simeq 1.0$.

The orbital energy is:
\begin{eqnarray*}
\label{eq:orb_e}
\lefteqn{E_{orb}=  -\frac{G (M_{\rm 1}+ M_{\rm new,1})(M_{\rm 2} + M_{\rm new,2})}{R_{\rm sep}}} \\
&  &  + \frac{1}{2} (M_{\rm 1} + M_{\rm new,1})V_{\rm 1}^2 +\frac{1}{2} (M_{\rm 2} + M_{\rm new,2})V_2^{\rm 2}
\end{eqnarray*}
where $R_{\rm sep}$, the distance between the progenitors' centres of mass,
and $V_{\rm 1}$ and $V_{\rm 2}$, the centre of mass velocites of each 
progenitor, are defined at the beginning of the encounter.
Most of our simulations are nearly parabolic, such that this term is 
close to zero.  

\begin{figure*}
\begin{center}
\subfigure[]{\resizebox{8.0cm}{!}{\includegraphics{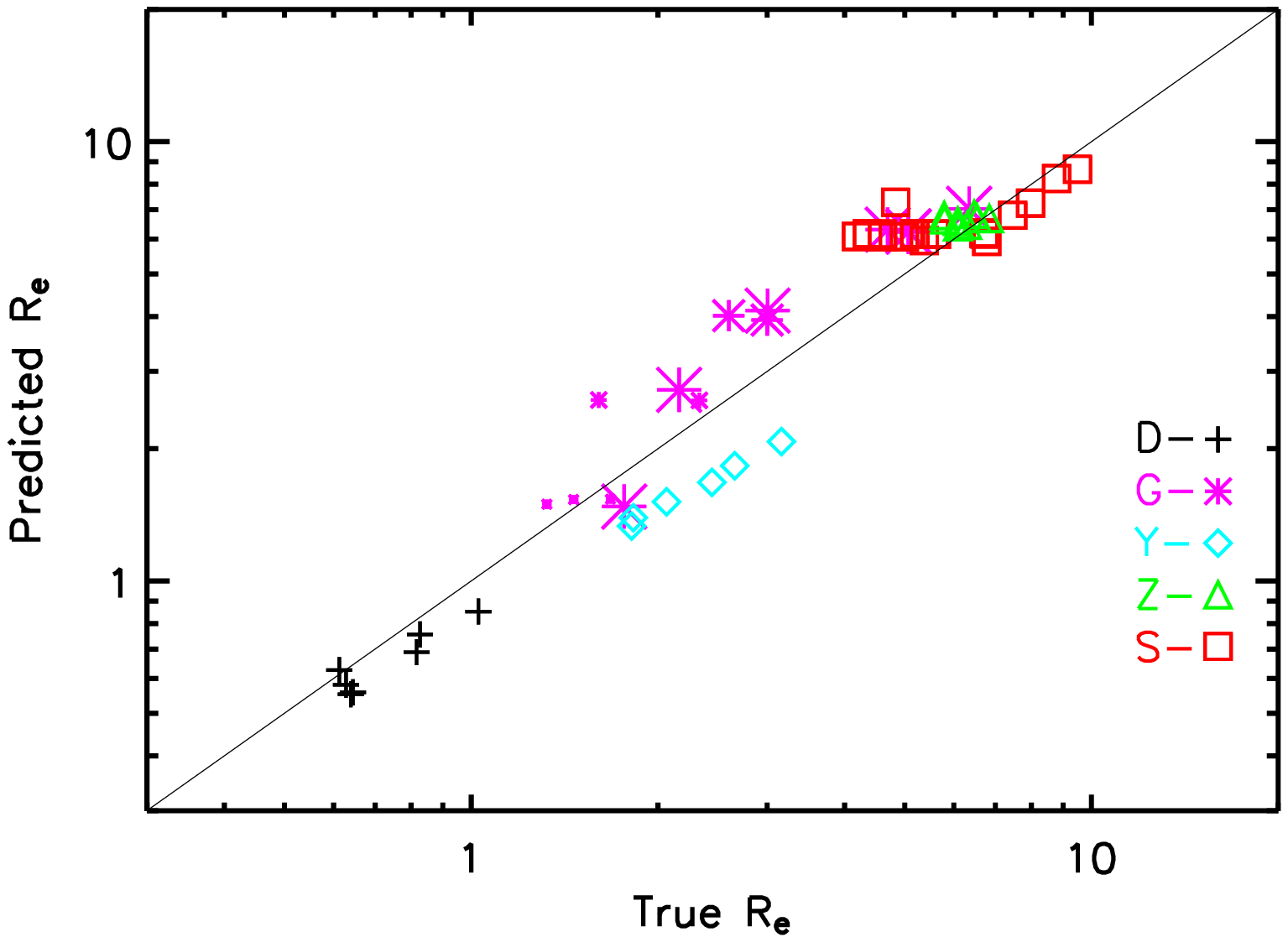}}}
\hspace{1.5cm}
\subfigure[]{\resizebox{8.0cm}{!}{\includegraphics{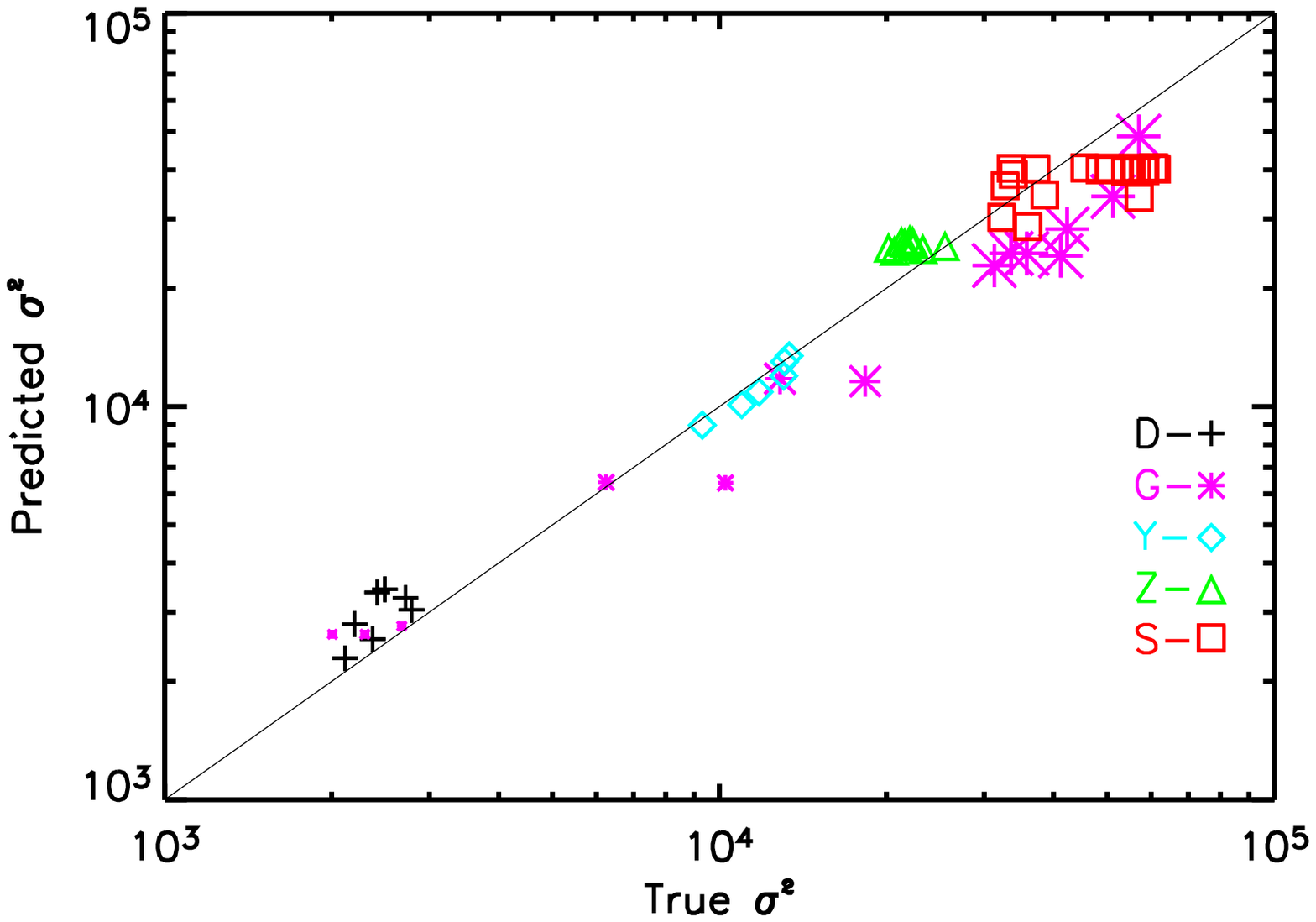}}}\\
\caption{The model predictions for radius (a) and velocity dispersion (b) 
versus the actual values. Predicted $R_{\rm e}$ and True $R_{\rm e}$ are the
predicted and simulated values of the three-dimensional stellar half-mass radius of the
remnants, respectively.  $\sigma$ is the line-of-sight velocity dispersion
inside the projected half-mass radius.  The scatter is $S=0.21$ for the radius 
prediction and $S=0.24$ for $\sigma^2$.}
\label{fig:major}
\end{center}
\end{figure*}

Our model predictions for the remnant radii are plotted 
against the actual stellar half-mass radii of the simulated merger 
remnants in Figure \ref{fig:major}. The overall scatter is $S=0.21$,
which is a significant improvement over the dissipationless model.
There is no obvious systematic error, and the fit is good both for the
wide range of progenitor properties and for the different choices of
orbital parameters. The spread for a particular progenitor type results
from differences in orbit and orientation.  For most cases 
this orbital spread is well fit by the model.  This is particularly evident
with the D and Y series, for which only orbit is varied and not orientation.
The spread in Sbc series is not quite as closely tracked by the model.  This
additional spread is the result of differing orientations.  Orientation is
not taken into account in the model for reasons described in \S \ref{ssec:Orbits}.
We note that the performance of the final model for radii is somewhat 
better than what might have been expected from the quality of the prediction of
radiative losses, shown in \ref{fig:rad_mod2}.  However, this is not as 
surprising if one remembers that radiative losses are only a correction on the
non-dissipative model.  Even if this correction is not exact, it always acts
in the proper direction and makes the remnants more compact.

\subsection{Velocity Dispersion}

\citet{NaabGas} have shown that the kinematics of merger remnants change
as a function of initial gas fraction.  When predicting the central velocity
dispersion of merger remnants we take gas fraction into account in two ways:
first by using the gas fraction dependent predicted radius, and second by 
adjusting the central dark matter fraction to account for the rearrangement of
gas and subsequent formation of new stars.  
In order to compute the stellar velocity dispersion of the merger remnants, we 
implement a virial relation of the type
\begin{equation}
\sigma^2 = C_{\rm vir} \frac{G M_{\rm dyn}}{R},
\end{equation}
where $M_{\rm dyn}$ is the dynamical mass of the system.  This mass includes
all of the stellar mass of the system and also a contribution from the
dark matter near the centre of the galaxy.
$C_{\rm vir}$ is a constant that varies slightly with galaxy structure and
also accounts for the conversion between the three-dimensional radius and
the projected, two-dimensional velocity dispersion.   
$R$ is a characteristic radius (e.g., the stellar half-mass radius).  
$\sigma$ is a line-of-sight velocity dispersion of stars inside the 
projected half-mass radius.  We measure this from the simulations by averaging 
over 50 random projections. 

Given the predicted values for $R_{\rm f}$ and the final stellar mass,
one may attempt to estimate $\sigma^2$.  
However, variations in central progenitor dark 
matter fractions and differing amounts of star formation result in 
significantly different ratios of $M_{\rm dyn}$ to the total stellar mass in 
the remnants.
Because we are dealing with an approximate virial relation of a system 
that actually is not completely self-gravitating, it is not obvious 
from first principles what the exact contribution to $M_{\rm dyn}$ from
dark matter should be.  We track the dark matter contribution by estimating
the dark-matter fraction for each remnant.  Therefore the uncertainty
above, concerning which dark matter mass to include, translates into an 
uncertainty in which radius to choose for defining a dark-matter fraction.  
Because we use the three-dimensional stellar half-mass radius in our virial relation, it would
be reasonable to choose a radius $\sim R_{\rm f}$.  
However, one can pick a range of radii and 
still achieve sensible results by making slight adjustments to $C_{\rm vir}$.
We find the best results when we focus on the dark matter fraction 
inside {\it half} of the three-dimensional stellar half-mass radius.  We
define the dark-matter fraction within a given radius by 
\begin{equation}
f_{\rm dm}=\frac{M_{\rm dm}}{(M_{\rm dm} + M_{\rm stars})},
\end{equation}
where $M_{\rm dm}$ and $M_{\rm stars}$ are the dark matter and stellar masses 
inside that radius, respectively.  Much of the variation in the remnant dark matter fraction,
$f_{\rm dm,f}$, is due to
the variation in initial dark-matter concentrations and baryon distributions.
However, another important effect is the tendency of new stars to form near the 
galaxy centre, causing mergers that form more stars to end up with lower
$f_{\rm dm,f}$ values.
We thus predict the final dark-matter fraction using the initial
dark matter masses and our model prediction for the mass in new stars,
\begin{equation}
f_{\rm dm,f}= \frac{M_{\rm dm,1} + M_{\rm dm,2}}{M_{\rm dm,1} + M_{\rm dm,2} + C_{\rm stars}(M_{\rm 1} + M_{\rm 2} + M_{\rm new})}.
\end{equation}
$M_{\rm dm,1}$ and $M_{\rm dm,2}$ are the dark matter masses inside half of the
three-dimensional stellar half-mass radii of the progenitors. $M_{\rm 1}$ and
$M_{\rm 2}$ are the stellar masses of the progenitors, and $M_{\rm new}$ is the
total mass of stars formed during the merger as predicted by equation \ref{eq:sf_imp}.
This expression simply assumes that the inner region of the remnant 
contains the same amount of dark matter as the sum of the inner regions 
of the progenitors, and that a fixed fraction, $C_{\rm stars}$, of the final 
stellar mass is inside one-half of the three-dimensional stellar half-mass radius. 
We find that the best fit to the simulated remnants is $C_{\rm stars} \simeq 0.35$.

Our modified virial relation becomes
\begin{equation}
\sigma^2 = C_{\rm vir} \frac{G (M_{\rm 1} + M_{\rm 2} + M_{\rm new})}{R_{\rm f} (1 - f_{\rm dm,f})}.
\end{equation}
The best-fitting value from our simulations is $C_{\rm vir} \simeq 0.30$. 
The model predictions for the stellar line-of-sight velocity dispersions
are shown in Figure \ref{fig:major}, in comparison with the simulated values. 
The scatter in $\sigma^2$ is $S=0.24$, and the overall performance 
is similar to that of the model for the final radii. There are no obvious
systematic errors, and the predictions properly capture the variations 
due to either progenitor properties or orbital parameters.

\section{Unequal Mass Mergers}
\label{sec:Mass}

\begin{figure*}
\begin{center}
\subfigure[]{\resizebox{8.0cm}{!}{\includegraphics{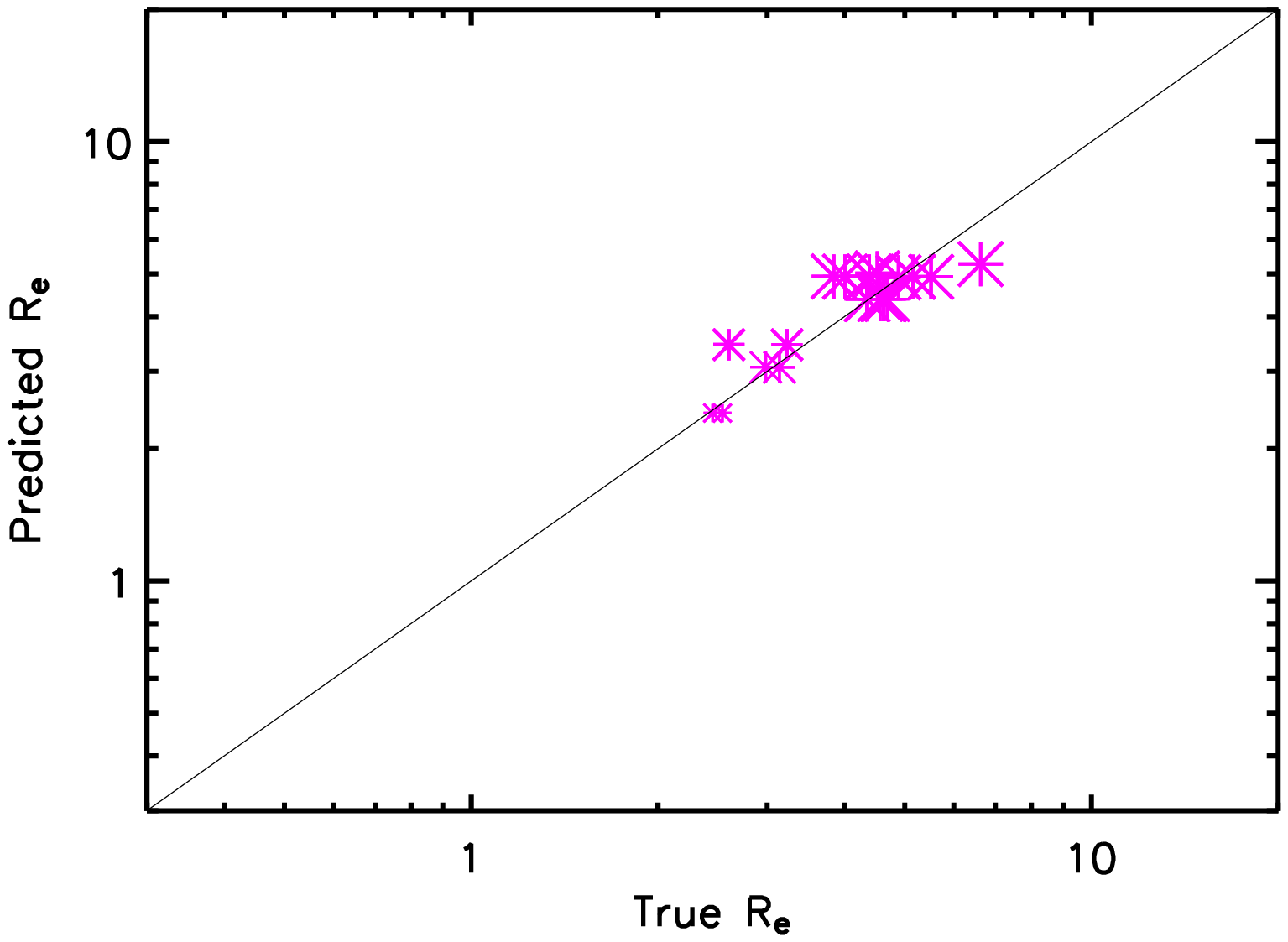}}}
\hspace{1.5cm}
\subfigure[]{\resizebox{8.0cm}{!}{\includegraphics{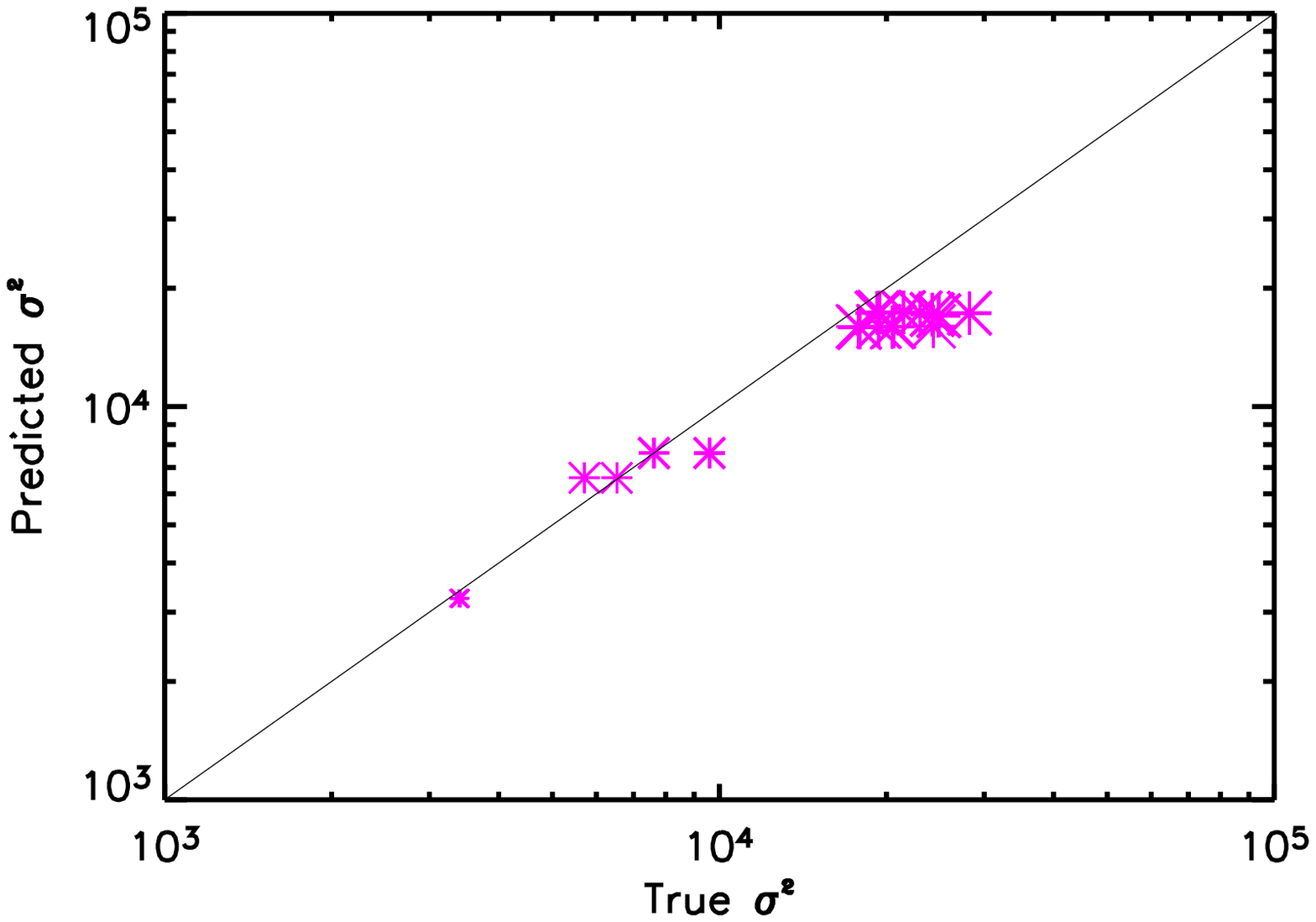}}}\\
\caption{Unequal Mass Mergers. The predicted values of our radius and
velocity dispersion models
versus the actual values for unequal mass mergers of G series galaxies. $S=0.13$ for radius, and $S=0.20$ for $\sigma^2$.}
\label{fig:unequal}
\end{center}
\end{figure*}

The cases studied so far were mergers of identical galaxies, and the model
was built using the simulations of these cases.  However,
most mergers in the universe are of unequal mass galaxies.  More recent 
simulations were run of unequal mass mergers \citep{minors}, and given the 
importance of unequal mass mergers in the real universe
we here examine the accuracy of our model for
different progenitor mass ratios.  For this study we used simulations of 
unequal mass mergers from the G series of galaxies \citep{minors}.  These 
mergers cover a range of mass ratios and a variety of orbital parameters.
For unequal mass mergers, the predictions of remnant properties remain
relatively accurate with
$S=0.13$ about the previously fit relation for radii, and $S=0.20$
for $\sigma^2$ (see Figure \ref{fig:unequal}).  Note that in some cases 
the spread does
not fall directly along the line.  This is the result of a number of 
simulations with the same progenitors and orbital parameters, but with 
different orientations of the progenitors with respect to the orbital plane.  
Variations due to orientation do not have clear 
enough systematics to be accounted for in the model.  We discuss this 
limitation in \S \ref{ssec:Orbits}.


In major mergers, which have mass ratios greater than 1:3, the gas disks of 
the galaxies are severely
disrupted, and we see large flows of gas toward the galactic centres.  
However, in minor mergers, which have mass ratios less than 1:3,
these flows are much less pronounced because the gas disk of the larger
progenitor is only modestly disrupted.  For example, central gas densities 
in a major
merger of two G3 galaxies reach values roughly four times that of a minor 
merger (1:6) of G3 and G1 galaxies on the same orbit.
Hence, dissipation is expected to be much less important in the centres 
of the big progenitors of minor mergers.  Our model captures this effect
through the impulse dependence of the radiation term.  For minor mergers where
 $M_{\rm big} \gg M_{\rm small}$, the radiation term becomes insignificant 
in comparison to the internal energy term.  It is also worth noting that the
the star formation equation (\ref{eq:sf_imp}) under predicts
star formation in the smaller progenitor in this extreme minor merger
regime.  However, this has little effect on the predictions of remnant
properties since the mass of new stars formed in a much smaller
progenitor is much smaller than the mass of the larger progenitor.

\section{Caveats}
\label{sec:Caveats}

\subsection{Feedback}
\label{ssec:feedback}

Arguably the largest uncertainty in the physics of our simulations is in the
prescription for feedback.  The merger simulations fit by our model all
use the same 
prescription for feedback with the same parameters, 
hence, it would be useful to vary these 
feedback parameters and examine the effect on our model.  
The feedback recipe
is characterized by two parameters: feedback efficiency and an equation of
state parameter, $n$, which sets the polytropic index.  The effective pressure is

\begin{equation}
P_{eff} \sim \rho^{1 + (n/2)}.
\end{equation}

We explore three equations of state, $n=0$, $n=1$, and $n=2$, where $n=0$ 
corresponds to an isothermal equation of state and $n=2$ results in a stiff 
equation of state.  The 
feedback efficiency determines how quickly the feedback energy is allowed to
thermalize.  Higher efficiencies result in a quicker dissipation of the 
feedback energy.  For each value of $n$, we examine two different values of 
efficiency.  As a lower limit,
we use a low efficiency value that gives just enough pressure to stabilize the
disk.  We also simulate cases of super-stable disks where the efficiency
is set to ten times that needed for disk stabilization.  
In table 2, labels of the feedback
parameter sets denote the values of $n$ and efficiency chosen.  Unstable cases
with even lower feedback 
efficiency were run but are not discussed here.  Thus, for consistency
our ``low'' efficiency cases are labeled ``med''(e.g. {\it n2med} means
$n=2$ and the feedback efficiency is low).  In general, with lower feedback, 
the maximum star formation rate is higher, but the starburst duration is shorter. 
For a more detailed description of our feedback model see \citet{Cox05, minors}.

  To examine the effects of feedback on our remnant model, we take the 
fiducial Sbc merger and simulate identical initial conditions with varied 
values of the feedback parameters. As expected, feedback has a significant 
effect on star formation.  The effect is twofold.  First, both increased 
feedback efficiency and, to a smaller extent, stiffer equations of state 
(higher n) 
decrease the total number of stars formed.  This can be a dramatic effect.  
As shown in Table \ref{tab:feedback} the total number of stars formed varies by
more than a factor of three.  Secondly, feedback can
change the radial distribution of stars formed.  A
stiffer equation of state will lead to more stars being formed at large 
radii \citep{Cox05}.

\begin{table}
\begin{center}
\caption{Effect of differing feedback parameters on remnant properties. 
All simulations in the table are variations of a fiducial Sbc merger.
$M_{\rm newstars}$
is the total mass in new stars formed during the 3 Gyr simulation, burst and
quiescent.  $R_{\rm f,1/2}$ is the stellar {\it three-dimensional half-mass radius} of the remnant.
$\sigma_{\rm f} $ is the velocity dispersion of the remnants measured inside of
the {\it projected half-mass radius} averaged over 50 random projections.}
\begin{tabular}{p{2.0cm}p{1.5cm}p{1.5cm}p{1.5cm}}
\hline
  Model & $M_{\rm newstars}$ &  $R_{\rm f,1/2}$ & $\sigma_{\rm f} $\\
   & ($10^{10} \msun$) &  (kpc) & (\kms)\\
\hline
\hline
n0med & 5.75 &  5.37 & 253\\
n0high & 2.68 & 5.72 & 218\\
\hline
n1med & 5.30 & 5.34 & 252 \\
n1high & 2.57 & 5.75 & 190\\
\hline
n2med(fiducial) & 5.03 & 5.62 & 225\\
n2high & 1.61 & 6.95  & 178 \\
\hline

\end{tabular}
\label{tab:feedback}
\end{center}
\end{table}

\begin{figure*}
\begin{center}
\subfigure[]{\resizebox{8.0cm}{!}{\includegraphics{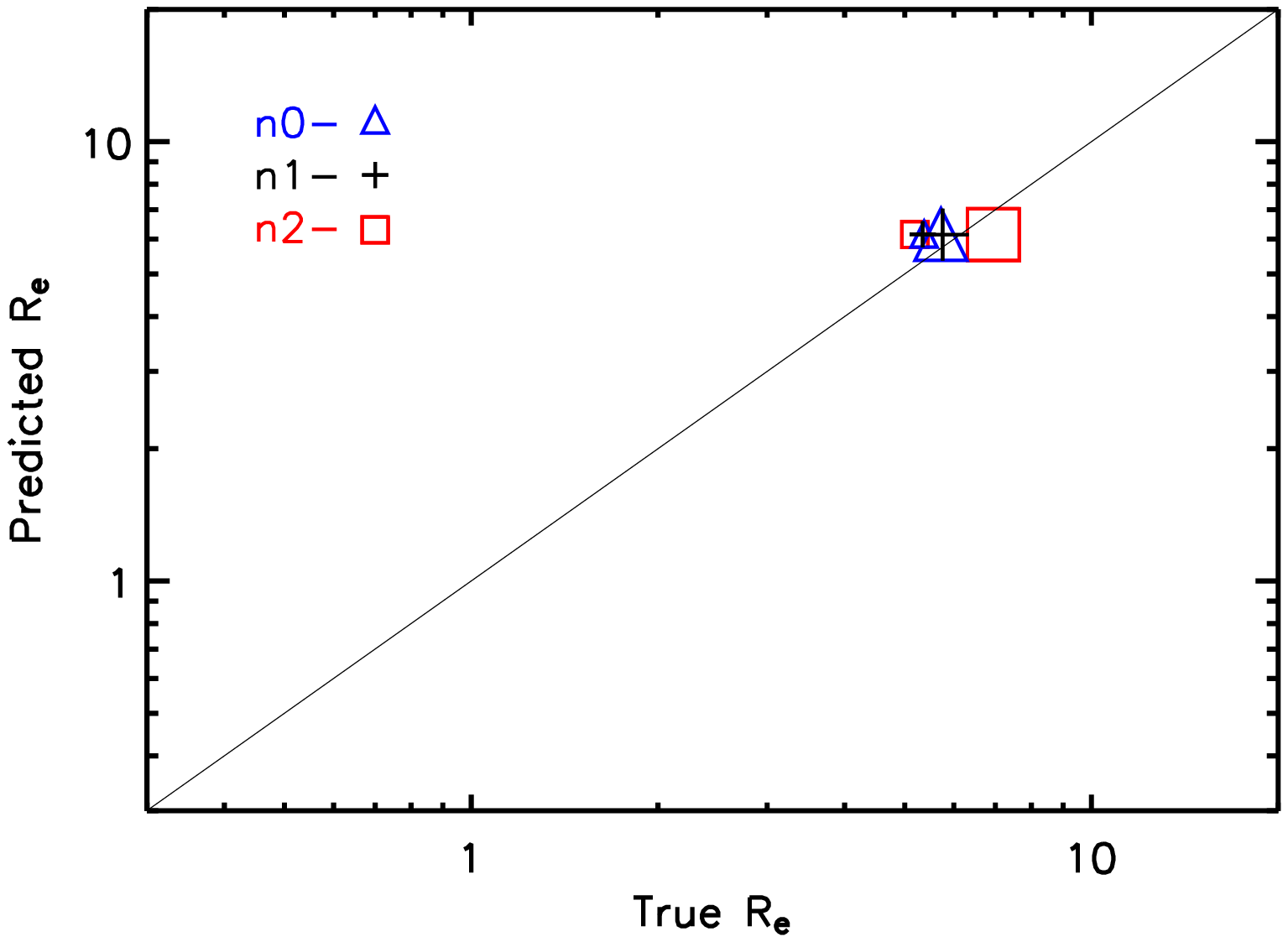}}}
\hspace{1.5cm}
\subfigure[]{\resizebox{8.0cm}{!}{\includegraphics{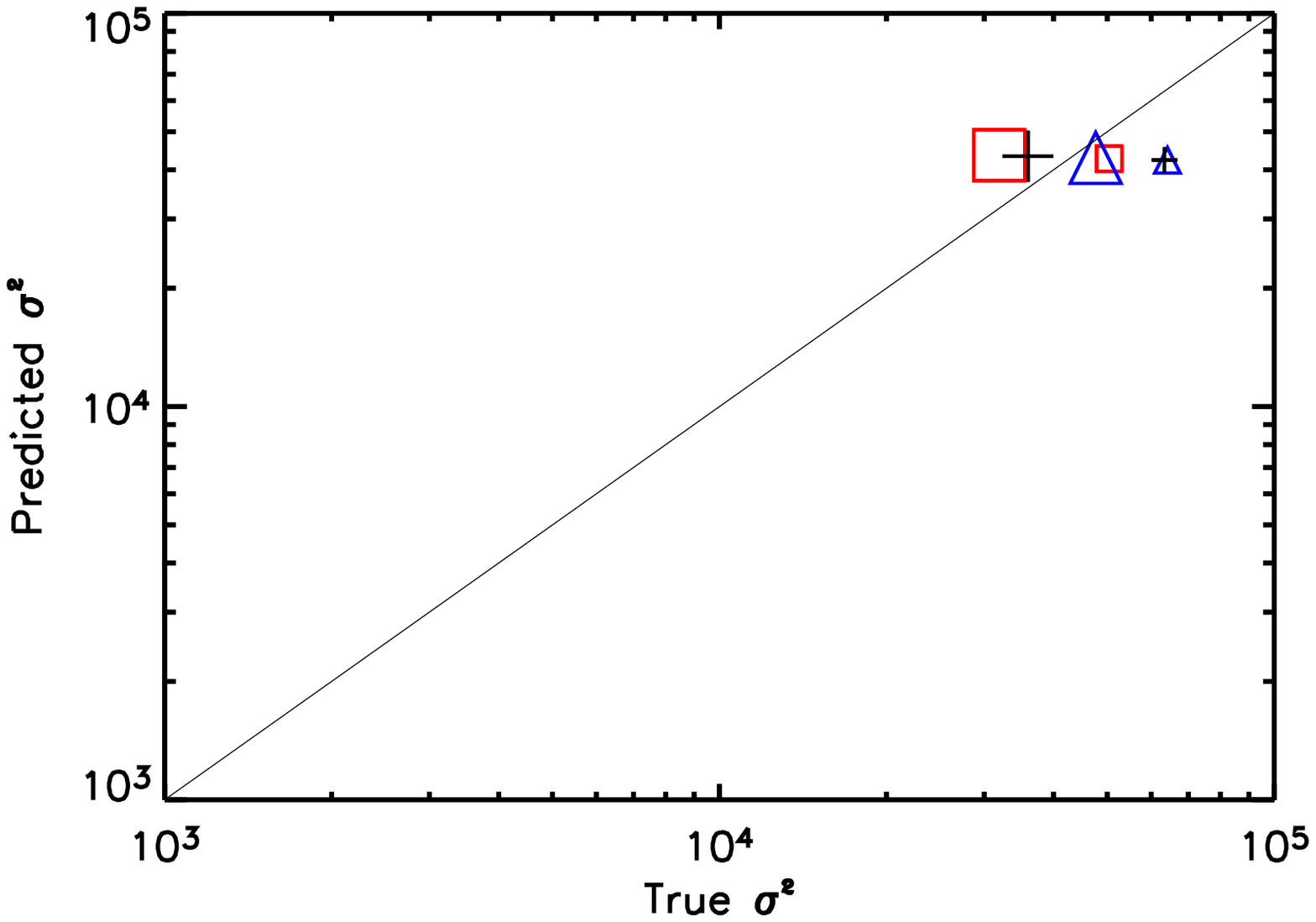}}}
\caption{Various feedback models. The predicted values of radius (a) and velocity dispersion (b) for all of the 
variations on the feedback model.  Large symbols represent high feedback efficiency,
small symbols represent low efficiency, and 
squares, crosses, and triangles represent n=2, n=1, and n=0 respectively.}
\label{fig:fb_mod}
\end{center}
\end{figure*}

  Both of these effects can act in concert to increase the size of the 
merger remnant.  However, the effect is weak enough that it takes both
high feedback efficiency and a stiff equation of state to significantly
alter the final half-mass radius (see Figure \ref{fig:fb_mod}).
  In our set of feedback models, the only
model for which the radius is significantly different than the others is 
{\it n2high}.  Even in this model the remnant radius is only a factor of $\sim 1.3$ 
larger than the most compact remnant. 
  Differences in radius and compactness will also affect the central velocity
dispersion.  Higher efficiency and stiffer feedback models produce somewhat
lower $\sigma$.  This effect is similar in magnitude to the effect on 
radius, with a factor of $\sim 1.4$ between the largest and smallest $\sigma$.
For velocity dispersion, our fiducial model lies near the centre
of the distribution (see Figure \ref{fig:fb_mod}).
These differences are roughly comparable to the scatter in our remnant model. 

While we have chosen
a specific feedback model as a fiducial, and used it to run most of our cases, 
we note that there are currently no theoretical or observational motivations 
for this choice.
Feedback in the real universe could resemble any of these models.
If feedback in the real universe is significantly different from our fiducial
case, then it is possible that the parameters of our model would require some
tuning.  $C_{\rm new}$ is the parameter most affected by differing feedback.  
The variations in star formation suggest that $C_{\rm new}$ could vary roughly
from $0.1$ to $0.35$.  It is also possible that the structural parameters
 $C_{\rm int}$, 
$C_{\rm vir}$, and $C_{\rm stars}$ would need to be adjusted.  
However, aside from
the mass of new stars, the relatively drastic changes in feedback produce
relatively minor changes in remnant properties.
This suggests that any tuning needed to match the real universe
would be small.  Further verification of the model will require testing
with independent feedback recipes.

\subsection{Orbits}
\label{ssec:Orbits}

There are several potential sources of uncertainty concerning our treatment 
of the initial orbits of the progenitors.  
The first regards the orbital energy term in our energy conservation equation.
\citet{Cole00} assume in their dissipationless model that some of the 
orbital energy is transferred to the outer dark matter halo through dynamical 
friction.  In order to estimate the energy transferred from the orbit
to the baryonic remnant, they assume that the two galaxies come into a 
circular orbit at a radius that equals the sum of their half-mass radii.  
However, the orbits in our simulations actually tend to quickly radialize 
after the first close passage. We therefore adopt the better approximation 
that all the orbital energy ends up in the baryonic remnant.
This too may be an over-simplification, as the energy deposited in the 
dark halo is likely to depend on the initial eccentricity of the orbit,
and may get larger for increasingly tangential orbits.
Unfortunately, our suite of simulations is not particularly well-suited to
exploring this dependence, as most of them start with initial orbital 
eccentricities close to unity. 
Our suite does contain three Sbc simulations and one Z simulation 
with $e=0.8$, and also one unequal mass merger, a G3-G2, with an 
initially circular orbit, $e=0$.  We find that these few cases
are well fit by our model as is.  However, given our limited sampling
of non-radial orbits, we keep in mind that our orbital-energy
term may require some modification.  On the other hand, we note that 
statistical studies of cosmological merger orbits suggest that they tend to be
rather radial \citep{Benson05, KandB06},  and that all attempts we made to 
include a term for energy transfer to the dark matter did not significantly
improve the model.  

A second potential concern is the initial orbital 
parameters we use.  Our simulations typically start
when the centre of one galaxy is near half of the virial radius
of the other galaxy, whereas the typical analysis of halo orbits in 
N-body simulations refer to orbits at the virial radius \citep{Benson05,KandB06}.  
Furthermore, our simulations involve the final entry into this 
radius, whereas in cosmological situations many of the halos enter and 
leave the virial radius more than once before merging.
One practical concern is how to convert between
orbits at $0.5\,R_{\rm vir}$ and at $R_{\rm vir}$.  Dynamical friction undoubtedly
induces some evolution in the orbital parameters as the galaxies fall between
$0.5\,R_{\rm vir}$ and $R_{\rm vir}$.
This effect should be evaluated using cosmological N-body simulations.

Finally, our model does not take into account the effect of the initial 
orientation of the progenitors with respect to the orbital plane.  
Our suite of simulations does include a number of 
different initial orientations, including both prograde and retrograde 
encounters.  Previous authors have observed systematic differences between
encounters with different orientations \citep{NaabGas, Barnes02}, but the 
effects on the remnants in our simulations do not 
show clear enough systematics for us to effectively characterize them.
The performance of the model could be improved by including
a term that depends on orientation, but the modest improvement did not 
seem to justify the inclusion of yet another parameter that would
increase both the complexity of the model and 
the risk that the model is over-fit. 

\subsection{Gas Fraction}
\label{ssec:gf}

Our merger suite primarily consists of disky progenitors with high gas 
fractions.  It is possible that we are missing some of the effects
that dominate the results of dry mergers of spheroids.  For example, 
\citet{Boylan06} find that more radial encounters of dry ellipticals 
result in larger remnants. However, we see no evidence for such a trend, 
not even in our lowest gas fraction simulations, the Z series, 
where $M_{gas}/M_{baryon}=0.1$.  In fact, for these low-gas cases, 
we see essentially no systematic trend of remnant size with orbit, 
and a generally smaller scatter in remnant size.  
We conclude that if we miss any effect of this sort, it is likely to be 
small.    Apparently, the dissipationless energy conservation equation
becomes a relatively accurate approximation in the limit of low gas 
fraction, and our model converges to the dissipationless model 
when the progenitors are gas-free.  

  Another limitation of our model is that it is unlikely to apply to extremely
gas-rich mergers, where $M_{\rm gas} \gg M_{stars}$.  For such cases the 
initial energy term should account for the size of the gas disk, whereas our 
model only takes into account the initial size of the stellar disks.  The 
highest gas fraction case that we include has a ratio of 
$M_{\rm gas}/M_{\rm stars} = 3.1$.  For the cases that we simulate, where
$M_{\rm gas} \lesssim M_{\rm stars}$ we find that the final sizes are 
relatively insensitive to the initial sizes of the gas disks.  Specifically,
our suite
includes cases where the gas disk scale length is equal to the stellar disk 
scale length  and cases where the gas disk scale length is three times the
stellar disk scale length.  Both of these cases are well fit by our model.

One final note is that our merger simulation suite does not attempt to 
model galaxies at higher redshift by systematically changing gas fraction
and concentration as has been done in other studies \citep{RobertsonFP}.  
Since progenitors are primarily modeled after low-redshift galaxies, this
is the regime where the model is most well-calibrated.  However, the model
is based on simple physical principles and is robust to systematic changes in 
gas and bulge fraction. We see no
obvious reason why it should not apply to binary mergers at higher redshifts.

\section{Conclusions}
\label{sec:Conclusions}

We have developed a simple toy model for the physical processes involved in wet 
mergers of galaxies, and have calibrated it using a suite of hydrodynamical
merger simulations.
This modeling helped us to gain a better understanding of these processes,
and provides a practical semi-analytic recipe for predicting 
post-merger galaxy properties in SAMs.

Crude models of this sort have been used by \citet{Cole00}, 
\citet{Galics03}, and \citet{Shen03}, but these models did not account for energy losses through
dissipative processes, and they have not been calibrated against 
realistic merger simulations.
Using a suite of merger simulations, we have demonstrated the key role of 
dissipative energy losses in determining the final radii and velocity 
dispersions of  merger remnants.
We found that the dissipative effects depend on the
initial orbits of the progenitors.  More violent, lower angular momentum orbits
create greater disturbances in the gas disks, which in turn radiate more energy
and produce more stars.  This orbital ``violence'' can be parameterized through
an impulse approximation for energy exchange between the orbital and internal
components during the first close pass of the encounter.  

We present a physically-motivated, simulation-calibrated model that is 
capable of predicting star formation,  central dark matter fraction,
remnant radius and remnant velocity dispersion, given the properties of
the progenitors and the initial orbital parameters of a merger.

The non-dissipative energy
conservation model often predicts radii that are off by a factor of 
$\sim 2-3$, and it does not reproduce the spread due to
orbital variations. Our model, which  
accounts for the dissipative energy losses, results in only $\sim 25\%$
errors in the predicted radius and velocity dispersion when a wide
variety of progenitor types is considered. For a given 
progenitor type, the error in remnant properties is reduced to  
$\sim 10\%$, indicating that our model correctly captures the variation 
of remnant properties due to merger orbit.  

Since we used the whole available simulation suite to calibrate our model,
via a few proportionality constants of order unity,
a proper evaluation of the model performance is yet to be pursued using
an independent suite of simulations.

\bibliographystyle{mn2e}
\bibliography{tj,matt,patriks}

\section{Appendix A: Impulse Approximation}
\label{sec:appA}

A number of previous researchers have studied and parameterized the transfer of
orbital to internal kinetic energy during galaxy encounters \citep{Richstone75, Dekel80, Aguilar85}.  \citet{BT} summarize these studies and present formulas
for impulse approximations valid in the tidal and radial cases.  
\citet{Aguilar85}
 find that the tidal approximation breaks down at approximately $5R_e$, but 
that a smooth interpolation between the two cases gives an approximate fit
to actual energy transfers:  
\begin{equation}
Tidal~ case:\\
\Delta E = \frac{2 G^2 M_{\rm 1} M_{\rm 2}^2 {\bar{r^2}}}{3R_{\rm peri}^4 V_{\rm peri}^2}
\end{equation}
\begin{equation}
Radial~ case:\\
\Delta E = \frac{3 G^2 M_{\rm 1} M_{\rm 2}^2}{3 V_{\rm peri}^2 a^2}
\end{equation}
$M_{\rm 1}$ and $M_{\rm 2}$ are the masses of the perturbed and perturbing galaxies 
respectively, and $R_{\rm peri}$ and $V_{\rm peri}$ are the pericentric distance and
velocity.  For the tidal case, ${\bar{r^2}}$ is the mean-square radius of the
perturbed galaxy.  A Plummer model, $\Phi = -GM/\sqrt{r^2 + a^2}$,
is assumed for the radial approximation.

  We measure the change in internal kinetic energy of each progenitor during
the first close encounter and define this as our ``impulse.''  In order
to compare with the above approximations we assume that the galaxies are
point masses and use values of $R_{\rm peri}$ and $V_{\rm peri}$ calculated
at the beginning of the simulations.
The impact parameters simulated in our suite all fall well within $5R_{\rm dm}$, 
as might be expected for {\it merger} orbits.  Consequently, the tidal 
approximation performs quite poorly at predicting the impulse.  Furthermore,
any impulse approximation assumes that the dynamical time of the perturbed
galaxy is much less than the time of the encounter.  However, if  
two galaxies are going to merge, then the encounter velocity is typically of the same order
as the internal velocity of the larger galaxy.  Therefore, it is not clear that
either impulse approximation above would apply.  We compared the radial 
approximation, and several similar functions that included a dependence on 
impact parameter, to the measured impulses and found that they did not
provide satisfactory fits.  

However, we do find that the measured impulse, during an encounter with orbital
parameters within our range of simulated values, is well-fit by the following
formula:
\begin{equation}
\Delta E = \frac{A G^2 M_{\rm 1,tot}^2 M_{\rm 2,tot}}{V_{\rm peri}^2 (R_{\rm peri}^2 
+ B~R_{\rm 1,tot}~R_{\rm peri} + C~R_{\rm 1,tot}^2)},
\end{equation}
where $M_{\rm 1,tot}$ and $M_{\rm 2,tot}$ are the total mass of the perturbed and 
perturbing galaxies respectively, baryons plus dark matter,
and $R_{\rm 1,tot}$ is the total half mass radius of the perturbed galaxy.
The parameters are set by best fit to the simulation results with
$A=1.6$, $B=1.0$, and $C=0.006$.

For a radial orbit, this formula vaguely resembles the radial approximation.  
However, the mass dependence of the formula is different.  In the radial 
impulse approximation $\Delta E \propto M_{\rm 1} M_{\rm 2}^2$, whereas in our 
formula $\Delta E \propto M_{\rm 1}^2 M_{\rm 2}$.  This suggests that the
impulse approximation is breaking down, that we cannot simply assume that
the potential of the perturbed galaxy is constant during the encounter.
 At intermediate impact parameters, where $R_{\rm peri}<R_{\rm 1,tot}$, 
the fitting formula falls off as $1/R_{\rm peri}$.  At larger impact parameters, 
$R_{peri}>R_{dm}$, it falls off as $1/R_{peri}^2$.
The parameter $C$
sets the cutoff radius at which the impulse ceases to increase with
decreasing impact parameter and approaches the constant radial case.
The fit of $C$ is somewhat tenuous since our simulations do not actually
probe the range of $R_{peri}^2 < C R_{dm}^2$.  However, these cases are 
so radial that
the probability of such encounters is low, and the effects of small changes
in $C$ are likely to be insignificant.  Our suite probes a range of roughly
$0.1 R_{\rm 1,tot} < R_{\rm peri} < 2 R_{\rm 1,tot}$ and 
$2 \sigma_{\rm prog} < V_{\rm peri} < 10 \sigma_{\rm prog}$, where $\sigma_{\rm prog}$ is the initial velocity dispersion of the larger progenitor.  The 
formula is plotted against
impulses measured from our simulations in Figure \ref{fig:impulse}.  This plot
includes both equal and unequal mass mergers.  The 
fit is valid over a range of progenitor mass distributions.  Specifically, the 
structure of the fitted galaxies varies widely with a variety of bulge fractions
and dark matter concentrations.

\section{Appendix B: Summary of the Model}
\label{sec:appB}

\begin{table}
\begin{center}
\caption{Definitions of the inputs used in the model.}
\begin{tabular}{p{2.0cm}|p{4.5cm}}
\hline
\hline
  Name & Definition\\
\hline
\hline
\underline{Inputs}\\
\hline
$M_{\rm tot,1}, M_{\rm tot,2}$ & The total masses (baryonic plus dark) of galaxies 1 and 2 respectively.\\
\hline
$R_{\rm tot,1}, R_{\rm tot,2}$ & The three-dimensional half-mass radii of the total masses (baryonic plus dark) of galaxies 1 and 2 respectively.\\
\hline
$V_{\rm peri}, R_{\rm peri}$ & The theoretical pericentric velocity and distance of the first encounter, defined when the galaxies are separated by $\sim R_{vir}.$\\
\hline
$K_{\rm tot,1}, K_{\rm tot,2}$ & The total kinetic energies of the galaxy/halo systems of galaxy 1 and 2 respectively. In our simulations $K_{\rm tot,1} \simeq 0.35~ G M_{\rm tot,1}^2/R_{\rm tot,1}$ with very little scatter.\\
\hline
$f_{\rm g,1}, f_{\rm g,2}$ & The gas fractions of galaxies 1 and 2, defined as (gas mass)/(total mass).\\  
\hline
$R_{\rm 1}, R_{\rm 2}$ & The stellar three-dimensional half-mass radii of galaxies 1 and 2.  \\
\hline
$M_{\rm 1}, M_{\rm 2}$ & The stellar masses of galaxies 1 and 2. \\
\hline
$M_{\rm dm,1}, M_{\rm dm,2}$ & The dark matter mass inside 1/2 the stellar three-dimensional half-mass radius of galaxies 1 and 2 respectively.\\
$M_{\rm stars,1}, M_{\rm stars,2}$ & The stellar mass inside 1/2 the stellar three-dimensional half-mass radius of galaxies 1 and 2 respectively.\\
\hline
\hline
\end{tabular}
\label{tab:input}
\end{center}
\end{table}

\begin{table}
\begin{center}
\caption{Definitions of the parameters used in the model.}
\begin{tabular}{p{2.0cm}|p{4.5cm}}
\hline
\hline
  Name/Value & Description\\
\hline
\hline
\underline{Parameters}\\
\hline
$A=1.6 $ & \multirow{3}{4.5cm}{Parameters fit to match the impulse model to the simulations. See Appendix A.}\\
$B=1.0$ & \\
$C=0.006$ & \\
\hline
$C_{\rm new}=0.3$ & Proportionality constant in star formation equation.  Determines mass of new stars.\\
\hline
$C_{\rm int}=0.5$ & Structural constant which sets relative weighting of internal energy.\\
\hline
$C_{\rm rad}=1.0$ & Constant which sets relative weighting of radiated energy.\\
\hline
$C_{\rm sig}= 0.30$ & Structural constant which sets proportionality of $\sigma^2$ to $M/R$ in remnants.\\
\hline
$C_{\rm stars}=0.35$ & Sets fraction of stars within 1/2 of the half-mass radius of the remnants.\\
\hline
\hline
\end{tabular}
\label{tab:param}
\end{center}
\end{table}

\begin{table}
\begin{center}
\caption{Definitions of the outputs used in the model.}
\begin{tabular}{p{2.0cm}|p{4.5cm}}
\hline
\hline
  Name & Definition\\
\hline
\hline
\underline{Outputs}\\
\hline
$M_{\rm f}$ & Mass of stars in the remnant (old + burst).\\
\hline
$R_{\rm f}$ & Stellar three-dimensional half-mass radius of remnant.\\
\hline
$f_{\rm dm,f}$ & The dark matter fraction inside 1/2 of $R_{\rm f}$\\
\hline
$\sigma_{\rm f}$ & The stellar velocity dispersion of the remnant inside the projected half-mass radius.\\
\hline
\hline
\end{tabular}
\label{tab:output}
\end{center}
\end{table}

\begin{figure*}
\begin{center}
\includegraphics[width=0.9\textwidth]{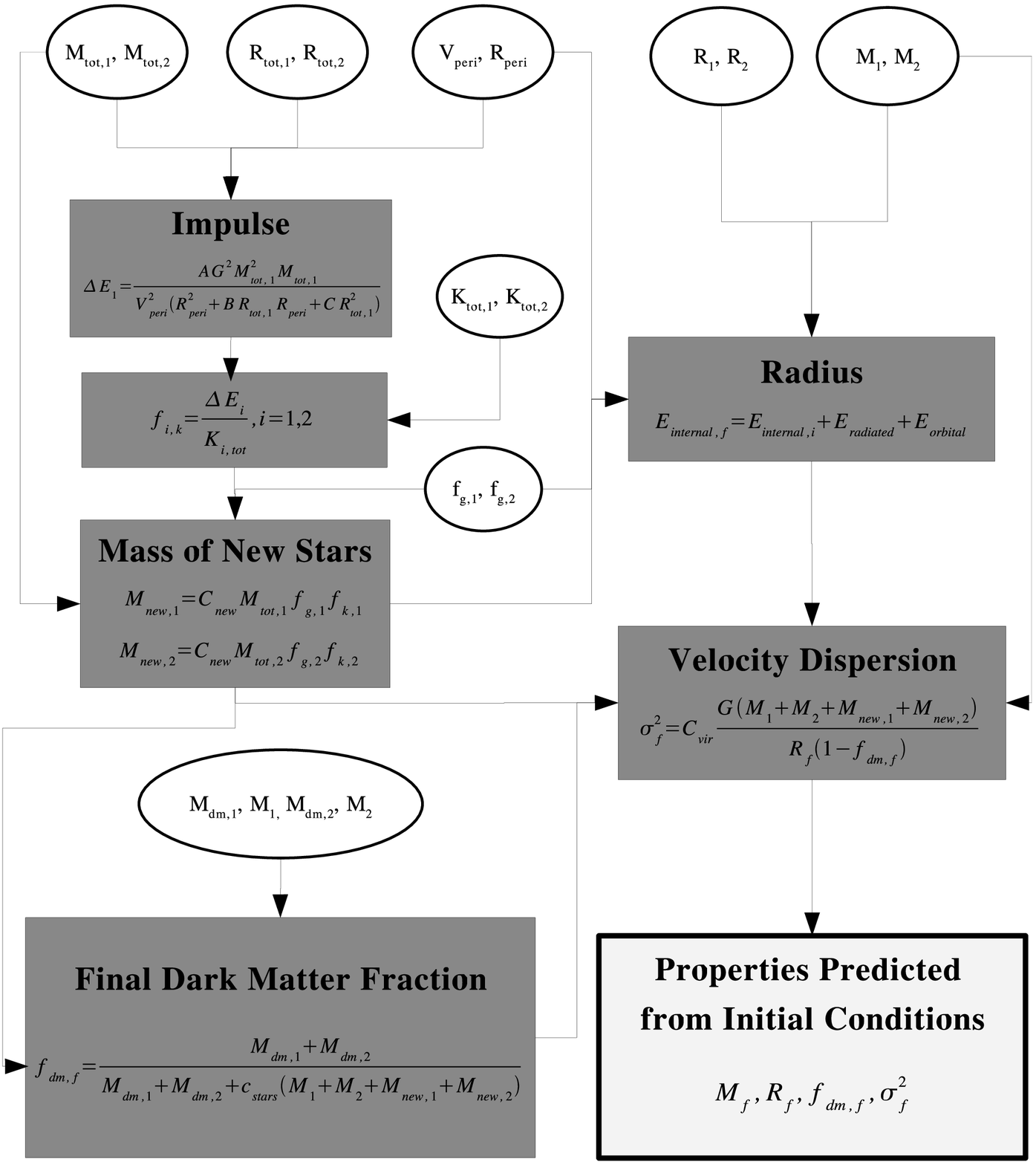}\\
\caption{Summary of the inputs (ovals), outputs (white rectangle), and equations (shaded rectangles) needed to implement the merger model.}
\label{fig:flowchart}
\end{center}
\end{figure*}

The model presented requires the integration of 
formulas and ideas found throughout the paper.  Hence, for pragmatic purposes,
we present a brief summary of the model for the reader who wishes to implement
it as a recipe within a semi-analytic model of galaxy formation.  
In Figure \ref{fig:flowchart} we illustrate the
outline of the model.  The relevant equations are included, except for the 
definitions of the energy terms in the conservation equation.  For these, we 
refer the reader to the list in \S \ref{ssec:diss_model}. Inputs, parameters,
and outputs are defined in Tables \ref{tab:input}, \ref{tab:param}, and \ref{tab:output}.

The general flow of the model is as follows: 
\begin{enumerate}
  \item[1.]{Use orbits and masses to calculate impulse.}
  \item[2.]{Use impulse and gas fraction to calculate new stars formed in the burst.}
  \item[3.]{Use mass of new stars and properties of progenitors to calculate radius.}
  \item[4.]{Use initial dark matter fraction and final mass of stars to calculate remnant central dark matter fraction.}
  \item[5.]{Use central dark matter fraction and radius to calculate velocity dispersion.}
\end{enumerate}

\end{document}